\documentclass[twocolumn,showpacs,preprintnumbers,amsmath,amssymb]{revtex4}

\usepackage{graphicx}
\usepackage{dcolumn}
\usepackage{bm}

\begin{document}

\noindent

\preprint{}

\title{A class of traveling-envelope solutions of free Schr\"odinger equation generated by Lorentz transformation}

\author{Agung Budiyono}

\affiliation{Institute for the Physical and Chemical Research, RIKEN, 2-1 Hirosawa, Wako-shi, Saitama 351-0198, Japan}

\date{\today}

\begin{abstract} 

We develop a class of traveling-envelope solutions of Schr\"odinger equation for a free particle  whose amplitude is moving with constant group velocity while keeping its shape undistorted. We show that solution with arbitrary finite group velocity is obtained by Lorentz boosting the solution with vanishing group velocity, if the quantum average energy $E$ and momentum $p$ are related to the rest-mass $m$ of the particle by Einstein formula $E^2/c^2-p^2=m^2c^2$. The wave function is spatially localized with finite-size support which is decreasing as the rest-mass and/or group velocity are increased. For a particle with vanishing rest-mass yet finite momentum, we show that the group and phase velocities are equal to the velocity of light and the wavelength is given by Einstein another formula $\lambda_P=h/p$. 
 
\end{abstract}

\maketitle

Let us consider the quantum dynamics of a single free particle of mass $\mu$ in one dimensional space denoted by $x$. The complex-valued wave function at time $t$, $\psi(x;t)$, is assumed to satisfy the following free Schr\"odinger equation: 
\begin{equation}
i\hbar\partial_t\psi(x;t)=-\frac{\hbar^2}{2\mu}\partial_x^2\psi(x;t).
\label{free Schroedinger equation}
\end{equation}
It is evident that the above Schr\"odinger equation is not Lorentz invariant, thus is usually not regarded as the proper equation describing a relativistic free particle \cite{Ryder book}. Despite of this fact, below we shall show that it possesses a class of traveling-envelope solutions which is generated by Lorentz transformation parameterized by $\mu$. We shall then argue that it quantum mechanically represents a free relativistically moving particle. 

First, given a pair of real-valued parameters $(k,v)$, let us consider the following class of wave functions:
\begin{equation}
\psi_{v}(x;t)=A_v\cos\big[\gamma k(x-vt)\big]\exp\Big[\frac{i}{\hbar}(\mu vx- \gamma E_0t)\Big],
\label{Lorentz wave function}
\end{equation}
where $\gamma=\gamma(v)$ is some dimension-less factor assumed to depend on $v$.  One can then check by direct substitution that the above wave function satisfies the free Schr\"odinger equation of Eq. (\ref{free Schroedinger equation}) if $E_0$ is related to $k$ and $v$ as follows:
\begin{equation}
E_0=\frac{\gamma\hbar^2k^2}{2\mu}+\frac{\mu}{2\gamma}v^2=\frac{\hbar^2 k^2}{2m}+\frac{1}{2}mv^2,
\label{rest-energy-wave number-mass}
\end{equation}
where we have defined $m$ as $m=\mu/\gamma$, whose physical meaning will be clarified later. Notice that the wave function of Eq. (\ref{Lorentz wave function}) is a traveling wave function with group velocity $v_g=v$. To be precise, since the phase velocity $v_p=\gamma E_0/(\mu v)=E_0/(mv)$ is in general not equal to the group velocity, below we refer to it as traveling-envelope wave function. We shall discuss the case when $v_p=v_g=v$ at the end of the paper. Notice also that in general it is not the eigenfunction of Hamiltonian operator, $\hat{H}={\hat p}^2/(2\mu)$, where $\hat{p}\equiv-i\hbar\partial_x$ is the momentum operator, except when $v=0$. $A_v$ is a normalization constant. Hence, to be normalizable with non-vanishing $A_v$, the wave function must possess only a finite-length spatial support. We shall first focus on the dynamical property of the traveling-envelope wave function and discuss the issue of its normalization afterward. 

To proceed, let us assume a functional form of $\gamma=\gamma(v)$ so that for vanishing group velocity, $v=0$, it reduces to unity: $\gamma(0)=1$. Then, for the case of $v=0$, the wave function of Eq. (\ref{Lorentz wave function}) becomes stationary
\begin{equation}
\psi_0(x;t)=A_0\cos(kx)\exp(-iE_0t/\hbar). 
\label{resting wave function}
\end{equation}
The above wave function satisfies the free Schr\"odinger equation when $E_0$ is given by $E_0=\hbar^2 k^2/(2m)$. One can then easily verify that the traveling-envelope wave function of arbitrary finite group velocity $v$ of Eq. (\ref{Lorentz wave function}) can be obtained back from the resting (stationary) wave function of Eq. (\ref{resting wave function}), up to normalization constant, by the following coordinate transformation:
\begin{eqnarray}
x\mapsto \gamma(x-vt),\hspace{4mm}
t\mapsto \gamma\Big(t-\frac{mv}{E_0}x\Big).
\label{aby transformation}
\end{eqnarray}

If one further chooses the following form of $\gamma=\gamma(v)$ and $E_0$: 
\begin{equation}
 \gamma(v)=(1-v^2/c^2)^{-1/2},\hspace{2mm}E_0=mc^2,
\label{Einstein relation}
\end{equation}
where $c$ is the velocity of light, then one obtains the Lorentz transformation \cite{Bohm book}
\begin{eqnarray}
x\mapsto \frac{x-vt}{\sqrt{1-v^2/c^2}},\hspace{4mm}
t\mapsto \frac{t-vx/c^2}{\sqrt{1-v^2/c^2}}.
\label{Lorentz transformation}
\end{eqnarray}
This is the main formal result of the present paper. Below we shall attempt a physical interpretation in term of relativistic dynamics of a free moving particle. Let us remark before proceeding that for sufficiently small group velocity $|v|\ll c$, Eq. (\ref{Lorentz wave function}) reduces into Eq. (21) of Ref. \cite{AgungPRA2}, up to small correction of the order $\sim o(v^2/c^2)$. Moreover combined with Eq. (\ref{rest-energy-wave number-mass}), the choice of $E_0$ in Eq. (\ref{Einstein relation}) restricts $k$ to depend on $v$ and $m$, $k=k(m,v)$. In particular, given $\mu$, one can choose a pair of $\{m,v\}$ so that $\mu=\gamma(v) m$. The chosen value of $\{m,v\}$ then determines $k$ through Eqs. (\ref{rest-energy-wave number-mass}) and (\ref{Einstein relation}). 

First, one can straightforwardly show that the quantum average momentum and energy over the traveling-envelope wave function of Eq. (\ref{Lorentz wave function}) are given by
\begin{eqnarray}
\langle\hat{p}\rangle\equiv\int dx\psi^*(x;t)\hat{p}\psi(x;t)=\gamma m v\equiv p,\hspace{10 mm}\nonumber\\
\langle\hat{H}\rangle\equiv\int dx\psi^*(x;t)\hat{H}\psi(x;t)=\gamma E_0=\gamma mc^2\equiv E,
\label{quantum average energy and momentum}
\end{eqnarray}
where we have used $E_0=mc^2$ in the lower equation. Hence, multiplying both sides of Eq. (\ref{rest-energy-wave number-mass}) with $\gamma$, one obtains
\begin{equation}
E=\gamma E_0=\frac{\hbar^2(\gamma k)^2}{2\mu}+\frac{\mu}{2}v^2=\gamma\Big(\frac{\hbar^2 k^2}{2m}+\frac{1}{2}mv^2\Big), 
\label{energy-wave number-mass}
\end{equation}
which can be interpreted as describing relation among different kind of quantum energies. In particular, the second term on the right hand side $K\equiv\mu v^2/2= \gamma mv^2/2$ should be interpreted as the contribution from the relativistic kinetic energy of a particle of mass $\mu=\gamma m$ moving with velocity $v$.  The physical meaning of the first term will be clarified as we proceed. The total quantum energy $E$ is thus always positive definite.

One can then use the quantum average energy $E=\gamma mc^2$ and momentum  $p=\gamma m v$ given in Eq. (\ref{quantum average energy and momentum}) to define the relativistic energy-momentum four vector (two vector, in our case of one space dimension).  In particular, from Eq. (\ref{Einstein relation}) and Eq. (\ref{quantum average energy and momentum}), one directly obtains the following relation: 
\begin{equation}
\frac{E^2}{c^2}-p^2=m^2c^2.
\label{Einstein energy-momentum-mass relation}
\end{equation}
This is the well-known and important energy-momentum relation developed in the theory of special relativity  \cite{Bohm book,Einstein energy-momentum}, now it is shown to be valid for quantum average energy-momentum. Thus, the above relation can be interpreted as the quantum version of the energy-momentum relation of special relativity. For the case of vanishing group velocity, $v=0$ so that $p=0$, one obtains the rest-energy $E=mc^2=E_0$. $m=\mu/\gamma$ thus must be interpreted as rest-mass. 

On the other hand, Eqs. (\ref{rest-energy-wave number-mass}) and (\ref{Einstein relation}) give 
\begin{equation}
\frac{\hbar^2 k^2}{2m}+\frac{1}{2}mv^2=mc^2. 
\label{wave number-velocity-mass}
\end{equation}
The above result is of course missing in the corresponding classical theory of special relativistic dynamics, thus is quantum mechanically inherent. In particular, for vanishing group velocity, $v=0$, one has 
\begin{equation}
\frac{\hbar^2 k^2}{2m}=mc^2=E_0. 
\label{internal energy}
\end{equation}
Since the right hand side of Eq. (\ref{internal energy}) is rest-energy of the theory of special relativity, then it is intuitive to call the first term on the right hand side of Eq. (\ref{energy-wave number-mass}), $I\equiv \gamma\hbar^2 k^2/(2m)$, as the internal energy. Equation (\ref{internal energy}) thus shows us that the relativistic rest-energy should quantum mechanically be interpreted as the internal energy when the group velocity is vanishing. In particular, it relates the particle-like property of rest-mass $m$ with the wave-like property of wave number $k$ of the amplitude when $v=0$ through the constant $\hbar/c$, a combination between the two constants of quantum mechanics and special relativity. Equation (\ref{internal energy}) has been conjectured by the author in the previous work \cite{AgungFP1}. Hence, starting from non-relativistic quantum theory, we can also predict the existence of rest-energy developed in the theory of special relativity.  

In contrast to the above results, in the prevailing approach to quantization, one usually starts from a classical relation between energy and momentum and infers the corresponding quantum mechanical relation by the standard quantization rule: namely replacing the energy and momentum in the classical relation with the corresponding differential operators: $E\mapsto \hat{E}= i\hbar\partial_t$ and $p\mapsto\hat{p}= -i\hbar\partial_x$, respectively \cite{Isham book}. In this paragraph, $E$ and $p$ refer to classical energy and momentum, respectively. Hence, for example, the Lorentz invariant Klein-Gordon equation is obtained by applying the above quantization rule to the relativistic energy-momentum relation of Eq. (\ref{Einstein energy-momentum-mass relation}) \cite{Ryder book}. Moreover, the quantization of the non-relativistic energy-momentum relation, $E=p^2/(2m)$, obtained from Eq. (\ref{Einstein energy-momentum-mass relation}) by neglecting the rest-energy, $mc^2$, and other small correction of the order $o(v^4/c^4)$ where $v$ is the velocity of the particle, gives the free Sch\"odinger equation of Eq. (\ref{free Schroedinger equation}) \cite{Isham book}. 

In the above standard scheme of quantization of a classical system, the plane wave plays a very crucial role to represent a free particle. This is encouraged by the fact that plane wave is the eigenfunction of momentum and energy operators, and inserting it into the Schr\"odinger equation and the Klein-Gordon equation will reproduce the corresponding classical relation between energy-momentum. In contrast to this, we argue in this paper that a free particle should be represented by a traveling-envelope wave function of Eq. (\ref{Lorentz wave function}). In particular, using the traveling-envelope wave function in the Schr\"odinger equation, one gets the relativistic energy-momentum relation. Moreover, the rest-energy is quantum mechanically interpreted as internal energy for vanishing group velocity shown in Eq. (\ref{internal energy}). 

Next, the relation of Eq. (\ref{wave number-velocity-mass}) shows that the wave number of the amplitude $k_v\equiv\gamma k$ depends on the group velocity and the rest-mass through
\begin{equation}
k_v=\frac{mc}{\hbar}\gamma(v)\sqrt{2-v^2/c^2}=\frac{\gamma(v)}{\lambda_C}\sqrt{2-v^2/c^2},
\label{wave number in mass and velocity}
\end{equation}
where $\lambda_C=\hbar/(mc)$ is the Compton wavelength. For vanishing group velocity $v=0$ one thus has $k_0=\sqrt{2}\lambda_C^{-1}$.  On the other hand, for the maximum possible group velocity $v\rightarrow c$, one obtains
\begin{equation}
k_v\rightarrow k_c= p/\hbar, 
\label{wave number momentum}
\end{equation}
where we  have used the upper equation of Eq. (\ref{quantum average energy and momentum}) for $v=c$. Since in this case $\gamma(v\rightarrow c)\rightarrow\infty$, to have a finite $\mu=\gamma m$ thus $p$ and also $k_c$, then $m$ and $k$  must be vanishing, $m\rightarrow 0$ and $k\rightarrow 0$. Moreover, in this limiting case of $v\rightarrow c$, the internal energy $I$ and the kinetic energy $K$ are both approaching half of the total energy
\begin{equation}
I=K\rightarrow E/2=pc/2, 
\label{internal energy for photon}
\end{equation}
where we have again used Eq. (\ref{quantum average energy and momentum}) and Eq. (\ref{wave number momentum}). 

Now let us discuss the problem of normalization of the traveling-envelope wave function of Eq. (\ref{Lorentz wave function}). In order to make the traveling-envelope wave function of Eq. (\ref{Lorentz wave function}) normalizable, $\int dx|\psi_v(x;t)|^2=1$, one must introduce a finite length of spatial support. Namely one assumes that the wave function takes non-vanishing value only within a bounded set of real numbers. It is natural to assume that the set is connected, hence the support should be a closed and bounded interval of real number. On the other hand, in Ref. \cite{AgungPRA2}, we have argued that the wave function can self-generate its own boundary through the idea of self-trapping \cite{AgungPRA1}. First, for $v=0$, one chooses a class of wave functions whose amplitude is given by  self-consistently solving the following pair of equations: 
\begin{equation}
|\psi(x)|^2=\frac{1}{Z}e^{-U(x)/T},\hspace{2mm}U(x)=-\frac{\hbar^2}{2m}\frac{\partial_x^2|\psi|}{|\psi|}, 
\label{most likely wave function}
\end{equation}
where $Z$ is a normalization constant and $T\ge 0$ is real-valued parameter. In the so-called pilot-wave \cite{Bohm-Hiley book} and Madelung fluid \cite{Madelung paper} interpretations of Schr\"odinger equation, the above defined $U(x)$ is called as quantum potential. Then the amplitude of the wave function in Eq. (\ref{Lorentz wave function}) for $v=0$ thus $\gamma=1$ can be shown as the solution of Eq. (\ref{most likely wave function}) in the limit of $T\rightarrow 0$ \cite{AgungPRA2}. In this limit, $U(x)$ is constant  inside a finite length of interval given by
\begin{equation}
\mathcal{M}_0=[-L_0,L_0]\hspace{2mm} \mbox{where}\hspace{2mm} L_0=\pi/(2k),
\end{equation}
and is infinite at the boundary points $x=\pm L_0$. Accordingly, the envelope of the wave function takes non-vanishing value only inside $\mathcal{M}_0$, and vanishing at the lower and upper boundary points of $\mathcal{M}_0$. In this sense, we said that the wave function is self-trapped by its own self-generated $U(x)$ \cite{AgungPRA2}. 

Then, as shown in Eq. (\ref{Lorentz wave function}), since  for the case of non-vanishing value of group velocity $v$ the wave number of the envelope changes as $k\mapsto k_v=\gamma k$, it is natural to assume that the spatial support for the case of group velocity $v$ at time $t$ is given by
\begin{equation}
\mathcal{M}_v=[vt-L_v,vt+L_v]\hspace{2mm} \mbox{where}\hspace{2mm} L_v=\pi/(2\gamma k). 
\label{self-trapping vs support}
\end{equation}
The envelope of the wave function is still vanishing at the boundary points of its support, which is now moving at the velocity $v$. Hence the length of the support is 
\begin{equation}
\Delta_v\equiv 2L_v=\frac{\pi}{\gamma(v) k(m,v)}=\frac{\pi\hbar}{mc}\frac{\sqrt{1-v^2/c^2}}{\sqrt{2-v^2/c^2}},
\label{support-wave number-velocity}
\end{equation}
where we have used Eq. (\ref{wave number in mass and velocity}). In general, the length of the support is thus determined by $m$ and $v$. In particular, it is decreasing as we increase $m$ and/or $v$. Accordingly, the normalization constant also depends on $m$ and $v$, and is given by $A_v=2\gamma k/\pi$. 

Let us express the length of the support in term of energy-momentum. Using Eq. (\ref{quantum average energy and momentum}) in Eq. (\ref{support-wave number-velocity}) one obtains
\begin{equation}
\Delta_v=\frac{\pi\hbar c}{E}\frac{1}{\sqrt{2-v^2/c^2}}=\frac{\pi\hbar v}{pc}\frac{1}{\sqrt{2-v^2/c^2}}. 
\label{energy and momentum vs support}
\end{equation}
Next, let us again discuss two limiting cases. First, for vanishing group velocity $v=0$ one has $\Delta_0=(\pi/\sqrt{2})\lambda_C$. On the other hand, for the case of  $v\rightarrow c$, one gets
\begin{equation}
\Delta_c\rightarrow\frac{h}{2p},
\label{length of a photon}
\end{equation}
where $h=2\pi\hbar$ is the Planck constant. Notice that the wavelength $\lambda_P$ of the amplitude of the traveling-envelope wave function is given as twice of its length of support. In the limit $v\rightarrow c$ one thus obtains
\begin{equation}
p\rightarrow\frac{h}{2\Delta_c}=\frac{h}{\lambda_P}, 
\label{wavelength of a photon}
\end{equation}
which is just Einstein's quantization formula for photon \cite{Einstein photon formula 1,Einstein photon formula 2}. The above result can of course be obtained directly from Eq. (\ref{wave number momentum}) by noticing $\lambda_P=2\pi/k_c$. 

Finally, since in the limit $v\rightarrow c$, one has $k_v=\gamma k\rightarrow p/\hbar$ and $E=\gamma E_0\rightarrow pc$, the traveling-envelope wave function of Eq. (\ref{Lorentz wave function}) for particle with vanishing rest-mass yet finite momentum $p$ is approaching
\begin{equation}
\psi_{(v\rightarrow c)}(x;t)\rightarrow\frac{2p}{\pi\hbar}\cos\Big[\frac{p}{\hbar}(x-ct)\Big]\exp\Big[\frac{ip}{\hbar}(x-ct)\Big],
\label{wave function of photon}
\end{equation}
where the support is given by $x\in\mathcal{M}_c=[ct-L_c,ct+L_c]$ with $L_c\rightarrow \pi\hbar/(2p)$.  In this case, one has $\mu=p/c$. Notice that the phase velocity and group velocity of the above wave function are both equal to $c$. Conversely, assuming that the phase velocity is equal to the group velocity, $v_p=v_g=v$, one gets $v=E_0/(mv)$. Recalling the fact that $E_0=mc^2$, one has $v_p=v_g=\pm c$. Moreover,  the wave function of particle with vanishing rest-mass of Eq. (\ref{wave function of photon}) now satisfies the Lorentz invariant Klein-Gordon equation without mass term (ordinary wave equation) 
\begin{equation}
\Box\psi=0, 
\end{equation}
where $\Box=-(1/c^2)\partial_t^2+\partial_x^2$ is D'Alembertian operator. 

To conclude, we have thus shown that even though the Schr\"odinger equation of Eq. (\ref{free Schroedinger equation}) is not invariant under Lorentz transformation, there is a class of solutions given in Eq. (\ref{Lorentz wave function}) which is generated by Lorentz transformation. The latter means that  the solution for arbitrary finite group velocity can be obtained by Lorentz boosting the solution of vanishing group velocity. This leads to the relation between the quantum average energy and momentum of Eq. (\ref{Einstein energy-momentum-mass relation}), which takes exactly the same form with the one developed in the theory of special relativity. It is therefore reasonable to regard the traveling-envelope wave function of Eq. (\ref{Lorentz wave function}) as describing a free particle moving relativistically with the velocity given by the group velocity of the wave function, and whose energy and momentum are given by the quantum average energy and momentum respectively. 

Furthermore, starting from Schr\"odinger equation, we also predict the existence of inherently quantum mechanical internal energy, which, for the case of vanishing group velocity, reduces into the rest-energy developed in the theory of special relativity. This energy is further argued to determine the length of the spatial support of the traveling-envelope wave function when it is at rest. In general, the length of the support of the traveling wave function of Eq. (\ref{Lorentz wave function}) is decreasing as we increase the rest-mass $m$ and/or the velocity $v$ of the particle. Since the spatial support provides the region where the quantum effect takes place, say through superposition, then one can say that the smaller the size of the spatial support, the more classical and particle-like the wave function is. These two facts are in accord with our intuitive definition of classical regime which is the regime of large energy \cite{AgungFP1}. Combined together, we have thus suggested that  the Sch\"odinger equation  contains, as a classical limit, the theory that describes the relativistic motion of a free particle. 

Moreover, in the limit of vanishing rest-mass $m$ yet with finite momentum $p$, we show that the group and phase velocities of the wave function are equal to the velocity of light, $c$. The relativistic mass is then given by $\mu=p/c$ and the wave length $\lambda_P$ of the corresponding traveling-envelope wave function is related to the momentum by the Einstein quantization formula for a photon $\lambda_P=p/h$. In this case, the traveling-envelope wave function satisfies the Lorentz invariant Klein-Gordon equation with vanishing mass term. 

At formal level, a point should be mentioned. As long as Eq. (\ref{rest-energy-wave number-mass}) is valid, Eq. (\ref{Lorentz wave function}) will satisfy the free Schr\"odinger equation for arbitrary choice of functional form of $\gamma$ and $E_0$, even if  they are not given by Eq. (\ref{Einstein relation}). Hence, the coordinate transformation of Eq. (\ref{aby transformation}) is in general not equal to the Lorentz transformation of Eq. (\ref{Lorentz transformation}). Accordingly, the energy-momentum relation of Eq. (\ref{Einstein energy-momentum-mass relation}) is valid only for the specific choice of  $\gamma=\gamma(v)$ and $E_0$ as given in Eq. (\ref{Einstein relation}) \cite{Svozil}. One should also note that the same choice of Eq. (\ref{Einstein relation}) gives the Einstein quantization formula for a particle with vanishing rest mass of Eq. (\ref{wavelength of a photon}). 

\begin{acknowledgments} 

This research is funded by the FPR program in RIKEN. 

\end{acknowledgments}

\end{document}